# Advancements in Terahertz Antenna Design

Sasmita Dash[1], and Amalendu Patnaik[2]
[1]University of Cyprus, Cyprus, [2]Indian Institute of Technology Roorkee, India
Email id: sdash@ec.iitr.ac.in, amalendu.patnaik@ece.iitr.ac.in

## Abstracts

The promising way to provide sufficient transmission capacity is by accessing transmission bands at higher carrier frequencies. This desire for higher carrier frequency or more bandwidth led the researchers to take advantage of the terahertz (THz) spectrum. The opportunity for large bandwidth in the THz band leads to the possibility of easy high data rate transmission. In spite of the advantages, the THz band suffers from large free space path loss. In the development of THz communication systems, the antenna is the most significant component. Especially, the focus is to design highly directive antennas, because it enhances the performance of the overall system by compensating the large path loss at THz and thus improves the signal-to-noise ratio. This chapter presents different types of THz antennas that include planar, reflectarray, horn antenna and lens antenna. Emphasis has been made to present the latest trend of designing THz antennas using carbon-based materials, such as graphene and carbon nanotubes. The performance of these antennas has been compared with that of traditional copper based THz antennas, by critically analyzing their properties. For completeness, a brief discussion on THz power sources has been made in this chapter. A comprehensive discussion on different fabrication techniques has been provided to appraise the reader of the general fabrication processes of THz components.

**Contents**

## List of Acronym

MIMO: Multiple input multiple output
THz: Terahertz
IR: Infrared
SPs: Surface plasmons
SPP: Surface plasmon polariton
Gbps: Gigabits per second
GHz: Gigahertz
NLOS: Non-line-of-sight
QCL: Quantum cascade laser
RTD: Resonant tunneling diode
OPM: Optical photoconductive materials
CW: Continuous wave
CMOS: Complementary metal-oxide-semiconductor
HBT: Heterojunction bipolar transistor
RF: Radio frequency
AMC: Artificial magnetic conductor
0-D: Zero-dimensional
1-D: One-dimensional
2-D: Two-dimensional
3-D: Three dimensional
LTCC: Low-temperature co-fired ceramic
SIW: Substrate integrated waveguide
SLM: Selective laser melting
CNT: Carbon nanotube
TM: Transverse magnetic
EDM: Electrical discharge machining
CNC: Computer numerical control
PCB: Printed circuit board
EM: Electromagnetic
FEM: Finite element method

# List of Symbols

$\lambda$: Wavelength

$d$: Distance

$A_{ma}$ : Molecular absorption loss

$A_s$ : Spreading loss

$\varepsilon_r$ : Relative permittivity

$\lambda_p$ : Plasmon wavelength

$\sigma_g$: Surface conductivity of graphene

$Z_S$: Surface impedance of graphene

$\hbar$: Reduced Plank's constant

$\omega$: Angular frequency

$\tau$ : Relaxation time

$E_f$: Fermi energy of graphene

$T$: Temperature

$e$: Electron charge

$K_B$: Boltzmann's constant

$v_f$: Fermi velocity

$\sigma_{CNT}$: Conductivity of CNT

$Z_{CNT}$: Surface impedance of CNT

$R_{CNT}$: Scattering resistance of CNT

$L_{CNT}$: Kinetic inductance of CNT

$m$ : Mass of the electron

$n$: Conduction electron density

$Z_{cu}$ : Surface impedance of copper

$R$ : Ohmic resistance

$L_i$ : Internal inductance

$L_k$ : Kinetic inductance

$L$ : Antenna length

$\lambda_0$: Free space wavelength

# 1. Introduction

The bandwidth requirements for wireless communications have increased rapidly over the last few decades. In order to tackle the situation, advanced modulation techniques have been applied to increase the spectral utilization efficiency [1]. This not only has increased the point-to-point data rates to a large extent but also enhanced the frequency reuse within a volume of space. However, the channel capacity upper limit is restricted by Shannon's formula, even with the use of multi-input multi-output (MIMO) strategy. Therefore, the only way to provide sufficient transmission capacity is by accessing transmission bands at higher carrier frequencies. This desire for higher carrier frequency or more bandwidth led the researchers to take advantage of the terahertz (THz) spectrum, that is, EM waves with frequencies ranging from 0.1 to 10 THz [2]. Besides this intrinsic advantage of high bandwidth, THz wireless communication has other advantages when compared with either microwave link or infrared (IR) based system, such as (i) more directional than microwave/millimeter links, (ii) secure, (iii) low attenuation compared to IR, and (iv) smaller scintillation effects compared to IR [3]. Because of these advantages, THz technology has grown dramatically over the last two decades and found its application in areas like communication, imaging, spectroscopy, biology/medicine, nondestructive evaluation, explosive detection, and radio astronomy. However, even today, these applications are not fully implemented due to the immature state of THz technology in terms of sources, detectors, antennas, and other essential components capable of working effectively in this frequency range. This chapter briefly discusses several THz source technique and THz antenna fabrication techniques for the possible realization of the THz wireless communication system.

In the last few years, several metal antennas and array structures, including planar antennas, horn antennas, reflectarray antennas, and lens antennas, have been designed for THz applications [18-48]. The implementation of metal antenna design in THz frequency remains a challenge. The dimension of the traditional metal antenna in this frequency range is in the order of micrometers. There lies a technological challenge in micro fabrications. Furthermore, at higher frequencies, the conductivity and skin depth of conventional metals decreases, leading to degradation of radiation efficiency and high propagation losses. The carbon-based nanomaterials (e.g., carbon nanotube and graphene) are emerging as promising materials for THz antenna designs [50-60] [65-78]. These antennas work on the principle of surface plasmons (SPs), which is a collective oscillation of electrons coupled to an electromagnetic (EM) field at a dielectric-metal interface [80]. The wave corresponding to the propagation of SPs is called the surface plasmon polariton (SPP) waves. The plasmonic antennas are such sub-wavelength structures on the surface of which SPs strongly confine. Properly designed plasmonic antennas can convert free-space EM radiation into SPs or can convert SPs into EM radiations. This paper discusses several THz antennas made up of graphene, carbon nanotube, and copper material. For completeness from the THz communication point of view, different THz sources and some fabrication techniques have also discussed in this chapter.

## 2. THz Communication

The THz band in the EM spectrum lies between the microwave and infrared frequencies. The THz (=$10^{12}$ Hz) radiation refers to EM radiation in the 0.1-10 THz range, *i.e.,* the region between the microwave and IR frequencies. The THz band had least explored by the researchers for a long time due to the unavailability of suitable sources. Currently, the research interest increased towards the THz range due to the availability of the THz sources and in the advancement in the laser, semiconductor, and optical photoconductive technology [81-107]. The different types of THz sources are discussed in the next section.

The THz EM spectrum has several benefits for wireless communications [3], [108]. The bandwidth requirements for wireless communications have increased rapidly over the last few decades. There is the only way to provide sufficient transmission capacity is by accessing transmission bands at higher carrier frequencies. THz spectrum opens the possibility for more bandwidth and high data rate transmission. Current wireless communications systems have bandwidths of a few GHz and data transmission rates up to 1 Gbps. The achievement of the data rates of 10 Gbps is comparatively difficult in the microwave band due to the narrow bandwidth. The data rate of 10–100 Gbps is realized by raising the carrier frequencies at 100–500 GHz [109]. The opportunity for large bandwidth in the THz band leads to the possibility of easy high data rate transmission [110]. THz communication is promising for wireless communications systems, particularly for the short-range indoor environment. However, even today, the THz communication system is not fully implemented due to the immature state of THz technology in terms of sources, detectors, antennas, and other basic components capable of working effectively in this frequency range.

In spite of the advantages, THz communication has a few limitations. In the THz band, the free space path loss is more significant than at lower frequencies [111]. This is the main reason to have less received power than the transmitted power. The free space path loss obtained from the Friis transmission equation,

$$\text{Free space path loss} = 20\log\left(\frac{4\pi d}{\lambda}\right) [\text{dB}] \qquad (1)$$

where $\lambda$ is the wavelength, and $d$ is the distance. According to the Friis equation, the THz band (small wavelength) leads to larger propagation loss. Furthermore, the THz signal suffers from both molecular absorption loss and spreading loss. The total path loss at THz band is the sum of molecular absorption loss $A_{ma}$ and spreading loss $A_s$. On account of the molecular absorption loss, several high attenuation levels are defined [31]. In the THz band, the spreading loss is 60 dB higher compared to the microwave band. Besides the path loss properties, the reflection properties in the THz band are different from the microwave band [112], [113]. The surface variations are on the order of THz wavelength (several hundred microns) for objects in indoor environments. Therefore, at the THz band, the indoor object surfaces are rough. We know from the scattering effect that the angle of reflection can be different from the angle of incidence for rough surface, and the receive antennas receive the signals from different spots, which leads to multi-path scattering [112]–[114]. At THz band, the multi-path scattering is significantly affected in both outdoor and indoor applications and makes the NLOS communication more challenging. Therefore, the THz band is merely appropriate for short-range communications where the range is in the order of a few tens of meters.

In the development of THz communication systems, the antenna is the most significant component. Antennas for THz communication have reported in the last few years. Especially, the focus is to design high gain antennas, because it enhances the performance of the overall system by compensating the substantial path loss at THz. The use of THz frequency allows for miniaturized antennas, which enables massive MIMO techniques for enhancement of spectral efficiency and directivity. In addition to high directive antennas, high power THz sources needed for the development of an effective THz wireless communication system.

## 3. THz Sources

In THz frequencies, the main problem is to get efficient sources. Different techniques for obtaining THz source found in the literature, namely quantum cascade laser, resonant tunneling diode, and optical photoconductive material techniques, complementary metal-oxide-semiconductor, heterojunction bipolar transistor [81-107].

Quantum cascade laser (QCL) may be a potential THz source of communication at the upper THz band, even if with the high atmospheric attenuation. The output power in QCLs increases when the signal frequency increases. The output power of 4.1 µW is achieved at 2.06 THz [81]. However, the QCL based THz source is not appropriate in the lower THz band. Efficiencies of QCL based sources are quite low at 1 THz region [82]. Several QCL based source is developed to attain more output power [81-87]. Using the QCL technique, the achieved maximum output power is more than 1 W [84]. In QCL, the tunneling of the electron through the barrier is possible without any extra energy, whereas, in resonant tunneling diode (RTD), the energy levels are quantized. The tunneling process in the RTD technique is high-speed. Thus, this technique is one of the efficient THz sources [88-92].

Recently, optical photoconductive materials (OPM) based techniques have been explored. OPM based THz sources are an efficient source in lower THz band and suitable substitutes to solid-state sources. This source also performs well in the upper THz band. The plasmonic electrode is now used for efficiency enhancement and achieves an efficiency of 7.5% [93]. Two different OPM based THz sources techniques are commonly used, such as pulsed technique and photo mixing technique. In both cases, photoconductive antennas are used for the emission of the THz radiation. In the pulsed technique, a femtosecond laser is used to obtain pulsed THz power, and in the photomixing technique, photo mixer is used to obtain continuous wave (CW) THz power.

Complementary metal-oxide-semiconductor (CMOS) based sources are used in the lower THz band. In the CMOS device, an active multiplier chain or a voltage-controlled oscillator is inserted [94]. The multiplier chain technique is widely used to achieve higher frequency output powers. This technique is used to attain 2.58 THz source and output power as 0.018 mW in [95]. In lower THz frequency between 0.288 THz to 0.498 THz, the output power from 22 µW to 2.6 mW is achieved in [96-100]. These results reveal that low operational frequency leads to high output power and low efficiency. Solid-state diodes are used to obtain the output power of 2.82 mW and 0.11 THz source in [101]. Higher output powers in higher frequency sources can be obtained using heterojunction bipolar transistor (HBT) technology. In [102-107], the output power from 12 µW to 10 mW, and the efficiency from 0.01% to 3.2% in frequency between 0.215 THz - 0.92 THz are measured. In solid-state devices, the output power reduces with the increase of THz frequency.

THz source techniques have several advantages and disadvantages. In view of the output power, CMOS is the best THz source in low THz frequencies. Above 0.3 THz, the output power in CMOS THz sources significantly reduce. In contrast, the QCL technique is not possible below 1 THz frequency. The OPM and RTD are the best THz source techniques in the 0.5 - 3 THz range. OPM THz source is the best suitable technique to achieve the best efficiency results. The CMOS and HBT THz sources also perform better at lower THz frequencies. Alternative to OPM, RTD is a suitable technique for THz bands. The insertion of plasmonic electrodes into the active photoconductive area improves the power and efficiency of THz radiation. The achievement of 1Tb/s wireless link using these efficient THz source techniques is not far away anymore.

## 4. THz Antennas

Copper is the well-known metal for the design of radio frequency (RF) and microwave antennas. The design of metal antennas in the RF and microwave frequency regime has thoroughly explored, and systematic antenna design methods exist [4]. However, the design of the copper metal antenna at THz frequency is entirely different from the classical microwave metal antenna. The classical microwave metal antennas excited through various feed lines such as microstrip, coaxial cable, and CPW, whereas the laser excitation through fiber or the air is seen in THz antennas. Metal THz antenna and microwave antenna have another difference, i.e., bias voltage. The design of the THz antenna needs bias voltage, but microwave antennas do not need any biasing. Furthermore, the fabrication of the metal THz antenna is more expensive and complicated than a metal microwave antenna. Over the past decades, several high directive metal THz antenna and array structures have been designed, including planar antenna and arrays [18]-[25], reflectarrays [26]-[36], lens antenna [37]-[43], horn antennas [44]-[48], carbon nanotube antennas [ 50-60 ], and graphene antennas [65-78 ].

### 4.1. Planar Antennas and Arrays

The unique features of the planar antenna, such as simple structure, low cost, and low-profile, motivates the antenna community to design different kinds of planar antennas to meet the future wireless communication requirements. However, the single-element antenna has a broad radiation pattern and small directivity values. In the wireless communication system, high directive antennas are needed for long-distance communication. Assemble of single-element in an electrical and geometrical configuration has the capability to provide sharp beam and high directivity (gain). This multi-element antenna is known as an array. A plethora of antenna arrays at MW frequencies are available in the literature for several applications [6-17].

Several planar antennas have been designed at THz frequency [18], [19]. One important issue has been addressed in most of these antennas, i.e., low antenna directivity. By assembling of 1000 single-element in the array, the antenna gain is enhanced to 31 dBi [20]. However, It is the sign of complex lossy networks. Planar antennas have designed at the low-THz frequency for beam scanning and high-gain applications [21], [22]. Nevertheless, the complexity in feeding and fabrication leads to more losses and shifts in the operational frequency band. The planar waveguide array antenna also has been developed at low THz frequency 0.12 THz [23]. The array provides 21.1 dBi gain of and 80% efficiency. A larger gain of 38 dBi and 43 dBi in 16×16 and 32×32 slot arrays are obtained in [24]. However, it is difficult to fabricate this multilayer waveguide slot array antenna, which has radiating waveguide and feeding network in the top layer and bottom layer, respectively. The fabrication process would be hard for a large and complex structure. The planar Yagi–Uda antenna is also promising for communication. The planar Yagi–Uda configuration exhibits high gain and low cross-polarization, which is useful for wireless communication. K. Han *et al.* reported a design of the Yagi-Uda antenna at low THz frequency 0.636 THz, which attained improved impedance matching using photomixer [25].

## 4.2. Reflectarrays

In the RF and microwave frequency region, reflectarray antennas have their own unique place because of getting the best performance of reflector and array antenna [26], [27]. In recent years, the reflectarray antenna design at THz frequency has been found in the literature [28 - 34]. The most crucial factor from these THz reflectarray antennas is element loss. Moreover, low-cost designs with satisfactory performance cannot be achieved from these works. The main reason for this is the occurring of conductor loss in the resonant element at THz frequency, which leads to a significant loss of power and tuning range of phase.

In [35], a dielectric reflectarray antenna at 220 GHz is reported. The reflectarray element consists of a dielectric block and a ground plane. The reflection phase is attained by controlling the block height. The 40×40 elements of dielectric reflectarray is fabricated using 3D printing technology and achieved 31.3 dBi gain and 27.6% aperture efficiency, and 20.9% 1-dB gain bandwidth. Another dielectric reflectarray antenna at 0.1 THz is fabricated by Objet Eden 350 polymer-jetting rapid prototyping 3-D printer [36] and 22.5 dB of gain is achieved.

## 4.3. Lens Antennas

Lens antenna arrays consist of an EM lens with the potential of energy focusing and a matching array with antenna elements placed in the focal region of the lens. An EM lens is a transmissive device, which has the ability to alter the EM rays propagation directions for the realization of beam collimation or energy focusing. EM lenses are employed by three different methods, such as i) the dielectric lenses, ii) the conventional planar lenses, and iii) the modern compact planar lenses. Generally, the conductor losses, feeding losses, and fabrication precision avoid scaling the high gain antennas from microwave to THz band. In this case, lens-based THz antennas are more suitable because of low losses. Several designs of lens antennas at THz frequency have been designed [37-43].

N. Llombart *et al*. reported a hemispherical silicon lens THz antenna at a low THz frequency of 545 GHz [37]. Mostly, silicon lens is used at a higher frequency for high-speed communication [38]. With the variation of the extension length behind the hemispherical position, the extended hemispherical system results in an antenna/lens system. In [39], the extended hemispherical lens with a double-slot antenna as the feed antenna and all the gold conducting layers deposited on a high resistivity silicon wafer. Extended hemispherical lens antennas are more suitable for beam scanning applications, especially with off-axis feeding [40], [41]. Bowtie THz antenna with a silicon lens and an artificial magnetic conductor (AMC) improves the radiation characteristics in terms of directivity, gain, efficiency, and the front-to-back ratio [42]. The bowtie-shaped antenna with silicon-lens at 1.05 THz attained directivity of 11.8 dB, realized gain of 11.6 dB, respectively, the radiation efficiency of 96%, aperture efficiency of 75.9%, the front-to-back ratio of 10.6 dB, and the maximum side lobe is 11.2 dB, and the bowtie-shaped antenna with AMC structure at 1.05 THz attained the directivity of 11.9 dB, realized gain of 10.7 dB, the radiation efficiency of 82%, aperture efficiency of 78%, front-to-back ratio of 9.1 dB and the maximum side-lobe level of 6.3 dB. H. Yi *et al*. reported 3-D printed dielectric lens based beam-scanning and High-gain THz antennas for THz communications, and radar applications. Here, dielectric lenses are used to increase directivity [43]. However, the implementation of antennas design in THz frequency remains a challenge. At THz band, the design of antenna demands a more complicated fabrication process, more cost, and more time.

### 4.4. Horn Antennas

Horn antenna is one of the widely used antennas due to its unique performances. It has its existence since the late 1800s. The horn is a flaring metal waveguide and commonly employed as a feed element. The performance of the radiator element depends on the type, direction, and amount of taper. It has widespread application because of its simple structure, easy excitation, wide bandwidth, more power capacity, and high gain.

In [44], a step-profiled pyramid horn antenna is investigated at low THz frequency 0.3 THz. The low-temperature co-fired ceramic (LTCC) multilayer substrate is used to achieve wide bandwidth and high gain. The stepped profile horn is structured through drilling cavities by increasing the size step by step on each layer using substrate integrated waveguide (SIW) technology. The horn antenna is fabricated with the process of LTCC multilayer and provides 100 GHz bandwidth and 18 dBi gain. The optimization of the step height and corrugation slot depth allows wide bandwidth and high gain. However, the fabrication complexity increases with the increase of substrate layers.

THz horn antennas are implemented using the metallic 3-D printing technologies in [45]. Antennas are printed using selective laser melting (SLM) technique on Cu–15Sn and sintering technique on 316L stainless steel. Horn antennas using the SLM technique on Cu–15Sn is developed at a low THz band from 0.11 THz to 0.32 THz. The antenna exhibit around 22 dBi of gain. 3-D printed metallic antennas are more simple and durable than nonmetallic 3-D printed antennas. As compared to the traditional technique for metallic horn antenna implementation, the cost of a 3-D printed antenna is low.

Different kind of horn antenna has been designed at THz [46-48]. The multiflare horn antenna is the best solution to achieve high directivity, and it provides directivity of 31.7 dBi at 1.9 THz [47]. However, at THz frequencies, the reduction of horn size leads to more difficulties in the fabrication, time, and cost.

## 4.5. CNT Antennas

Since the discovery of CNT in 1991, carbon nanotubes (CNTs) have attracted more attention among the researchers. In the last decade, the use of CNT in numerous fields found. Specifically, many antennas using CNT have designed at THz frequency due to its significant electrical properties. The current on a carbon nanotube antenna using a Fourier transform technique is investigated by G. W. Hanson [50]. The current on a carbon nanotube antenna is almost similar compared to current on copper antennas. For nanoscale radius, the CNT antenna has low losses than the copper antenna.

CNTs categorized as single-walled carbon nanotube (SWCNT) or multiwalled carbon nanotube (MWCNT). The theoretical study of the SWCNT dipole antenna at THz frequency is found in [51]. Wavenumber-domain integral equation is formulated by combining the Boltzmann transport equation and Maxwell's equations for obtaining the distribution of current. Furthermore, from this wavenumber-domain current, the radiation properties of the SWCNT THz antenna are obtained. The Numerical result of SWCNT THz antennas provides broad bandwidth and higher efficiency. SWCNT THz antennas have merits in terms of miniaturization, directivity, biocompatibility, and output power compared to metal photoconductive THz antennas. CNT antenna provides slow wave propagation, high input impedance, and low radiation efficiency [52- 55]. In the THz frequency, the CNT dipole antenna resonates at Length $L \approx \lambda_p/2$, $\lambda_p$ is the plasmon wavelength [55]. Above THz frequency range, strongly damped current resonance found, because of interband transitions in the optical frequency regime. The potential of CNTs as THz antennas of both the receiving and transmitting types has established in [56]. CNT antenna does not behave in the same manner as a nanowire antenna due to the variation in inductance [52]. The inductance of the CNT antenna is $10^4$ times of the nanowire antenna. This brings a significant difference in the performance of CNT and metal nanowire antennas. The behavior of the CNT antenna and metal antenna is entirely different due to their intrinsic high kinetic inductance and quantum capacitance [53]. The high kinetic inductance of CNT reduces the size of the antenna as well as reduces the antenna radiation efficiency. However, the radiation resistance and efficiency of the CNT dipole antenna in the THz band is low. Several CNT THz antenna designs have been designed to achieve the enhancement of the antenna efficiency using bundle structure [57-59]. In these antennas, the efficiency enhancement of the CNT antenna achieved at a low THz band by using a bundle of SWCNT. In the bundled SWCNT structure, SWCNT is surrounded by two dielectric jackets. The inner jacket is dielectric foam with relative permittivity $\varepsilon_{r1}$ close to unity ($\varepsilon_{r1} \approx 1$), and the outer jacket is a metamaterial layer with relative permittivity $\varepsilon_{r2}$ much less than unity ($0 < \varepsilon_{r2} \ll 1$). Radiation resistance and radiation efficiency of bundled SWCNT are increased by controlling the permittivity of the metamaterial jacket. The radiation efficiency can also be enhanced using MWCNT [60]. The radiation resistance and radiation efficiency increase with the number of layers and the frequency. Compared with a single MWCNT, a bundle achieves a significant improvement in radiation resistance and radiation efficiency. However, recently the unique properties of graphene material at THz frequency open an exciting scenario in THz antenna application.

## 4.6. Graphene Antennas

The latest addition of carbon allotropes family, graphene, is widely considered as the mother for the carbon allotropes of other dimensionalities, such as fullerene (0-D), carbon nanotubes (1-D), graphite (3-D). Recently, graphene attracted significant attention in various research fields due to its unique EM, mechanical, electrical, and thermal properties. Notably, the propagation of surface plasmon polariton (SPP) in graphene in the THz spectrum, which exhibits robust wave localization, moderate losses, and the tunable property through electrical/magnetic bias or chemical doping. Graphene plasmons are more easily tunable by changing the doping level via chemical or electrostatic gating in both single layer and bilayer graphene structure [63], [64]. Recently, the research interest is increasing for the realization of graphene antenna design at THz frequencies. Due to SPP propagation at THz, graphene enables plasmonic antennas at THz, whereas metal antenna made of noble metals such as gold, silver shows plasmonic behavior at optical frequencies. Plasmonic antenna resonates at the sub-wavelength scale with a high near field and strong coupling between localized sources and far-field radiation.

Graphene is used as an antenna radiator at the THz frequency range for the first time in 2012 [65]. In this work, the propagation properties of transverse-magnetic (TM) SPP in graphene has been used to model graphene patch antennas in the THz band. A THz CW photomixer is placed in the middle of the graphene patch to achieve the radiation. In addition to the high miniaturization, the graphene-based THz antenna performs better than metal implementations in terms of return loss and radiation efficiency. The radiation efficiency of the graphene antenna increases when the graphene chemical potential increases. Graphene plasmonic antenna enables high miniaturization and high directivity as compared to the metal antenna in the THz band [66]. Graphene-based antennas work at a much lower frequency than classical metallic antennas of the same size [67], [68]. Moreover, the performance of the graphene antenna at THz enhances by tuning the conductivity of graphene using an electric field effect. The performance merits of the graphene THz antenna are its high directivity, high miniaturization, stable impedance, and frequency reconfiguration [69]. Bilayer graphene provides dual-band reconfiguration with stable impedance, which avoids the need for a lossy and complex reconfigurable antenna [70]. Several works have further studied the capabilities of graphene in antenna design at THz frequency [ 71-78 ].

Although, several reflectarray and transmitarray antennas have been designed using conventional metal microstrip patches, dipoles, and dielectric resonator antennas, where metals are used as the electrical conductor in the reflectarray elements, the graphene reflectarray antenna provides a unique performance compared to metal reflectarray. Graphene reflective cells for THz reflectarray and graphene THz reflectarray antenna based on square graphene patches are designed at 1.3 THz [74], [75]. Graphene is used to control the phase of reflectarray at THz frequencies dynamically [75]. The plasmonic propagation supported by the graphene element allows drastically reduced inter element spacing and excellent array performance in terms of bandwidth, cross-polarization, and design simplification. Compared to a gold reflectarray, graphene reflectarrays exhibits better bandwidth, slightly lower cross-polarization, and easier design. Since the conductivity of graphene can be dynamically controlled using electric field effect, graphene reflectarrays enable reconfiguration using electric field effect via DC bias voltage.

MIMO technique in the wireless communications system is well-known for increase spectral efficiency. However, the size of the antenna and the separation between antennas are major obstacles for increasing the MIMO scale. Graphene antennas have the potential to reduce the size of antenna size and separation between antennas [77]. The spectral efficiency increased by the states of the graphene antenna array. The radiation patterns of graphene Yagi–Uda antenna easily reconfigures by controlling the properties of each graphene element [77]. The spectral efficiency of the graphene antenna based MIMO system has a higher spectral efficiency than the conventional MIMO systems based on metallic antennas. Generally, the THz channel has the characteristics of higher propagation loss and extra molecular absorption loss. Thus, the capacity drops in the THz band. Due to the unique properties of graphene, it can suit to design the reconfigurable directional antennas for THz communications.

## 5. Promising material for THz antenna

The materials so far used for antenna design in THz frequency are conventional metal copper, CNT, and graphene. However, the selection of promising material for THz antenna design is important for the THz community. This section provides the answer to this query by critically analyzing their properties at THz.

Graphene is a 2-D single layer of $sp^2$-bonded carbon atoms arranged in a hexagonal lattice. Graphene has an electron mobility of $2\times10^5$ cm$^2$V$^{-1}$s$^{-1}$ [61] and a current density of $10^9$ A/cm [62]. The surface conductivity of graphene consists of contributions from both intraband transition and interband transition. Based on Kubo formalism, the intraband conductivity dominates over the interband conductivity in the low-THz range, given by [79]

$$\sigma_g(\omega, E_f, \tau, T) = -j \frac{e^2 \kappa_B T}{\pi \hbar^2 (\omega - j\tau^{-1})} \left[ \frac{E_f}{\kappa_B T} + 2\ln\left(e^{-E_f/\kappa_B T} + 1\right) \right] \quad (2)$$

where $\omega$ is the angular frequency, $E_f$ is the Fermi energy, $\tau$ is the relaxation time, $T$ is the temperature, $j$ is the imaginary unit, $e$ is the electron charge, $\hbar$ is the reduced Planck's constant, and $k_B$ is the Boltzmann constant. The graphene layer behaves as a constant resistance in series with inductive reactance that increases with increasing frequency. The surface impedance of graphene shows the highly inductive nature of graphene surface conductivity at low THz frequencies [67]. Another important property of graphene at THz is the propagation of the SPP wave. Due to the two-dimensional nature of graphene, the SPP wave strongly confines at sub-wavelength scales. Surface plasmons in graphene layer exhibit unique properties of low losses, strong confinement, and high tunability. The surface conductivity of graphene plays a major role in determining the resonance of the graphene plasmonic antenna.

A CNT is formed by rolling a graphene sheet. CNT has an electron mobility of $8\times10^4$ cm$^2$V$^{-1}$s$^{-1}$ and a current density of $10^9$ A/cm [49]. Plasmonic waves in CNT can propagate at THz frequency. Due to the curvature effect, more plasmonic losses occur in CNT with less tunability as compared to graphene plasmonic material at THz frequency. CNT's conductivity consists of intraband conductivity and interband conductivity. In the THz frequency range, intraband

conductivity is more significant. The intraband conductivity of CNTs of the small radius expressed as [53]

$$\sigma_{CNT} = -j\frac{2e^2 v_f}{\pi^2 \hbar r(\omega - j\tau^{-1})} \quad (3)$$

where $e$ is the electronic charge, $r$ is the radius of CNT, $v_f$ is Fermi velocity ($v_f \approx 9.71 \times 10^5$ m/s for CNT), $\tau$ is the electron relaxation time, $\omega$ is the angular frequency, $\hbar$ is the reduced Plank's constant. The surface impedance of CNT calculated as

$$Z_{CNT} = \frac{1}{2\pi r \sigma_{CNT}} = \frac{j\pi\hbar(\omega - j\tau^{-1})}{4e^2 v_f} = \frac{\pi\hbar\tau^{-1}}{4e^2 v_f} + j\omega\frac{\pi\hbar}{4e^2 v_f} = R_{CNT} + j\omega L_{CNT} \quad (4)$$

where $R_{CNT}$ and $L_{CNT}$ are overall scattering resistance and kinetic inductance of CNT, respectively. The kinetic inductance increases with an increase in frequency. On the other hand, scattering resistance is independent of frequency. Therefore, scattering resistance of CNT remains constant with frequency in the low THz range.

Copper is an excellent conductor of electricity and the most widely used material in RF and microwave electronics. CNT has an electron mobility of 32 cm$^2$V$^{-1}$s$^{-1}$ and a current density of $10^6$ A/cm. Using Drude theory, the surface impedance and conductivity of copper in the THz frequency region expressed as [5]

$$Z_{cu} = \sqrt{\frac{j\omega\mu_0}{\sigma_D^{cu} + j\omega\varepsilon_0}} \quad \text{and} \quad \sigma_D^{cu} = \frac{ne^2\tau}{m(1+j\omega\tau)} \quad (5)$$

where $\omega$ is the angular frequency, $\tau$ is the electron relaxation time, $e$ is the electron charge, $m$ is the mass of the electron, and $n$ is the conduction electron density. Surface impedance of copper wire for small radius $r$ expressed as $Z_{cu}=R+j\omega(L_i+L_k)$, where $R$ is the ohmic resistance, $L_i$ is the internal inductance, and $L_k$ is the kinetic inductance. Kinetic inductance is much more than internal inductance and less than the ohmic resistance at low THz frequencies (where $\omega\tau \ll 1$). At 6.45 THz, $\omega\tau = 1$ for copper. Therefore, ohmic resistance is the dominant contribution to the surface impedance of copper wire below 6.45 THz frequency. The conductivity of graphene and CNT are more than copper [119]. The conductivity of graphene is higher than that of CNT. Therefore, graphene is the best conductor than CNT and copper in THz frequency.

In order to compare the performance of THz antennas made up of these carbon-based materials with that of the traditional copper metal, a rigorous analysis has done. In the first phase of this analysis, the physical size of the antenna is constant, whereas, in the second phase, the frequency is constant [66], [67]. The motive is to find the best material for the THz antenna design with superior performance.

In the first phase, a center-fed THz dipole of length ($L$) 71 µm (arbitrary dimension for obtaining low THz operational frequency) considered as the candidate antenna for comparison of the radiation performance for these three materials [67]. Antennas placed over silicon dioxide ($\varepsilon_r$= 3.9) substrate in all three cases. The antennas validated using Ansys HFSS, a finite element method (FEM) based EM solver. In the simulation, the graphene dipole antenna is modeled as the planar structure of the graphene sheet, whereas CNT and copper dipole antennas are modeled as a cylindrical structure of graphene and copper material respectively. The modeled structure of graphene, CNT, and copper dipole antenna is shown in Fig. 1. It found that the graphene antenna resonates at 0.81 THz, whereas CNT and the copper antenna resonate at 1.42 THz and 1.90 THz, respectively, which is illustrated in Fig. 2. From these results, it can be noted that the graphene antenna resonates at the lowest frequency compared to CNT and copper antenna.

Figure 1.THz dipole antennas. (a) Graphene, (b) CNT, and (c) Copper. © Springer Nature

Figure 2. S11 parameter and the far-field pattern of THz antenna of the same length (a) graphene, (b) CNT, and (c) copper. © Springer Nature

The radiation pattern of these dipole antennas at their resonant frequencies is shown in the inset of Fig. 2. Although their radiation patterns are identical, they differ in their directivities. It has been observed that the graphene antenna has higher directivity than the CNT antenna, and the directivity of the CNT antenna is higher than that of the copper antenna at low THz frequency range.

In the second phase, the performances of copper, graphene, and CNT dipoles with the same resonant frequency of 1 THz are analyzed [66]. At 1 THz frequency, the length of graphene THz antenna is 68 μm (= $\lambda_0/4.4$), whereas the length of CNT and copper THz antenna is 99 μm (= $\lambda_0/3$) and 139 μm (= $\lambda_0/2$) respectively. So, the graphene THz antenna achieves high miniaturization compared to CNT and copper THz antenna. The S11 parameters of these three THz antennas are shown in Fig. 3 along with their radiation patterns in the inset. This analysis also shows similar radiation patterns for three antennas with graphene THz antenna having maximum directivity. At 1 THz frequency, graphene, CNT, and copper dipole antennas have directivities of 4.27 dBi, 3.02 dBi, and 2.26 dBi, respectively. The performance of graphene, CNT, and copper antenna is summarized in Table.1. Although these results are shown for dipole antennas, it has been found that the results are equally applicable to other varieties of antennas and hence general in nature.

Figure 3. S11 parameter and far-field pattern of THz antennas at the same frequency 1 THz (a) graphene, (b) CNT, and (c) copper. © Wiley

Table 1: Performance of Graphene, CNT and Copper THz Antenna

The above analysis reveals that (i) Graphene is highly inductive and characterized by SPP at THz. Owing to excellent electronic properties and the propagation of TM SPP waves at THz band, graphene THz antennas exhibit higher directivity and higher miniaturization than copper and CNT THz antennas. (ii) The kinetic inductance of copper is much less than the ohmic resistance, and CNT has large kinetic inductance at THz band. Due to the kinetic inductive effect, CNT supports slow-wave propagation. (iii) At THz, owing to the support of slow-wave propagation and larger conductivity than copper, CNT THz antennas provide high miniaturization and larger directivity than copper THz antennas.

## 6. Fabrication of THz Antennas

Due to the persistent development of fabrication technologies, several new technologies are now ready to meet the processing requirements of THz. 3D printing or Additive manufacturing [115-118] is a method, which is used for the construction of 3D objects by printing them layer through the layer, primarily based on a 3D digital model. 3D printing is appropriate for structural components of large dimensions and complicated shapes. In 3D printing tools, the precision can attain up to 0.01mm. However, there is a need for sintering of the powder metallurgy parts after being printed. In the sintering process, the deformation of high temperature and shrinkage rate exist. The fabrication of high-precision parts is needed using the machining technique. The need for high precision processing of different electronic devices is now possible by several techniques such as low-temperature co-fired ceramic(LTCC), electrical discharge machining (EDM), computer numerical control (CNC) machining, and printed circuit board (PCB).

The fabrications of THz antennas using new materials are also pivotal. The latest carbon-based nanomaterial graphene has high current density and electron mobility than conventional metal copper. The conductivity of graphene and CNT are more than copper [119]. Recently, several researchers demonstrated in other contexts that graphene over hexagonal boron nitride leads to significant improvement [120] [121]. Due to its properties, graphene has promising potential for the development of effective electronic devices. The tunability behavior of graphene conductivity also enables the design of reconfigurable antennas. The research on THz antennas is now more useful and meaningful using new fabrication technologies and new nanomaterials.

High conductive graphene nanomaterial easily integrates with different substrate materials [122[ [123]. Graphene conductive ink is promising for printed electronics due to its unique mechanical and electrical properties. Recently, a few RF and microwave antennas are fabricated using graphene ink. An RFID dipole antenna is fabricated using graphene ink on the foam substrate by the rolling compression technique [122]. Another RFID antenna is fabricated using graphene ink by doctor-blading technique [124]. In this technique, graphene ink spread on the substrate using a mechanical mask, and the doctor blade is used for flattening and controlling the thickness. Graphene ink is used as a conductor in antenna-electronics interconnection fabrication using the 3D direct-write dispensing method [125]. In this case, the graphene antenna is fabricated using graphene ink on a 100% cotton fabric using 3D-printing technology. In [126], the CNT antenna is printed on the substrate and followed by cutting by the milling machine. The transfer technique also used to laminate the CNT on the substrate [127]. The fabrication of graphene, CNT, and metal antenna of the meander line dipole type is demonstrated using the direct ink-injecting technique at RF and microwave frequency [128]. The ink-injecting technology may have the potential for the fabrication of antenna for wireless communication applications. The fabrication of the graphene and CNT THz antenna is not yet fully explored. There is still a big room for improvement in the fabrication of graphene THz antennas, but the outlook is certainly promising, and this could have a significant impact soon.

## 7. Conclusion

In the design of any wireless communication system, the antenna is the most significant component.  In the THz wireless system, especially, the focus is to design high directive antennas, because it enhances the performance of the overall system by compensating the large path loss that occurs at THz frequency and thus improves the signal-to-noise ratio. This chapter discussed different THz antennas, made up of traditional metals, and carbon-based materials.

Generally, conventional metal copper is the most commonly used metal in the RF and microwave frequency range. However, the design of the THz copper antenna faces many challenges at THz frequency. Conductivity and skin depth of conventional copper metal are decreases with the increase of frequency. Low conductivity of copper leads to degradation of radiation efficiency of the copper antenna, and small skin depth of copper leads to high propagation losses at THz frequencies than microwave frequency. The materials other than conventional copper are explored to overcome these difficulties of the metal antenna at the THz band. A critical comparison of the performance of antennas made up of copper, carbon nanotube and graphene material shows that the antenna made up of graphene shows better performance in terms of radiation efficiency, directivity, and miniaturization.

Due to the unavailability of suitable THz sources and antennas, the THz band had least explored by the researchers for a long time. Currently, the research interest increased towards the THz range due to the availability of the THz sources due to advancement in the laser and semiconductor technology. This chapter discussed the different THz sources and fabrication techniques for the THz antenna for practical realization THz wireless communication system.